\documentstyle[12pt]{article}
\begin{document}
\setlength{\baselineskip}{20pt}
\begin{flushright}
RUP-5-95 \\
ITP-SU-95/01
\end{flushright}
\vspace{1cm}
\begin{center}
\Large{Polarization Effects in Chargino Production \\
at High Energy $\gamma\gamma$ Colliders} \\
\vspace{1cm}
\large{Masayuki Koike} \\
\normalsize{\it Department of Physics, Rikkyo University, Tokyo 171, Japan} \\
\large{Toshihiko Nonaka}\\
\normalsize{\it Department of Physics, Rikkyo University, Tokyo 171, Japan} \\
\large{Tadashi Kon}\\
\normalsize{\it Faculty Engineering, Seikei University, Tokyo 180, Japan}\\

\vspace*{1cm}

ABSTRACT
\end{center}
We investigate the chargino production process
$\gamma\gamma\rightarrow\tilde{W}^{+}\tilde{W}^{-}$
at high energy $\gamma\gamma$
colliders in the framework of the minimal supersymmetric standard model
(MSSM). Here the high energy $\gamma$ beams are obtained by the backward
Compton scattering of the laser flush by the electron in the basic linear
TeV $ee$ colliders. We consider the polarization of the laser photons as
well as the electron beams. Appropriate beam polarization could be
effective to enhance the cross section and for us to extract the signal from
the
dominant background $\gamma\gamma\rightarrow{W}^{+}{W}^{-}$.

\newpage
We call generally a class of the laws of nature standard model(SM).
The SM explains almost perfectly interactions between elementaly particles
at energy scale less than about 100 GeV.
However, it is known that the gauge hierarchy problem\cite{A} exists in the
model.
Now, we know, there are supersymmetric (SUSY) models \cite{mssm}
which could solve the gauge hierarchy problem.
The supersymmetry is a symmetry between fermions and bosons and
 the quadratic divergence can be cancelled out owing to
both contributions of bosons and fermions  . However,we should note that in the
exact SUSY limit fermions and bosons
must be degenerate in mass and there appears
 to be no evidence in nature for such a situation.
Therefore, in order to apply supersymmetry to
particle physics, we must consider models in which supersymmetry is broken.
In this case masses of SUSY partners must be less than about 1 TeV in order to
solve the hierarchy problem.

In this paper, we investigate
$\gamma\gamma\rightarrow\tilde{W}^{+}_{1}\tilde{W}^{-}_{1}$,
where $\gamma$ should be polarized. Here $\tilde{W}_{1}$ is the
lighter chargino
which is one of mixing states of the wino $\tilde{W}$ and the charged
higgsino $\tilde{h}$. We should note that the lighter chargino
$\tilde{W}_{1}$ could be the lightest charged SUSY particles in the
minimal SUSY standard model (MSSM). So there is a possibility that
$\tilde{W}_{1}$ would be discovered first at some high energy colliders.

The possibility for realization of $e\gamma$ and $\gamma
\gamma$ colliders have been discussed in detail by Ginzburg et al. \cite{gg}.
Here the high energy photon beams will be obtained by the backward
Compton scattering of the laser flush by one of electron beam in the
basic linear $ee$ colliders.
We consider polarized circular photons characterized by a four vector,
$$\epsilon=\frac{1}{\sqrt{2}}(0,\xi_2,-i,0), $$
where $\xi_2$ denotes the Stokes parameter of the  circular polarized photon
beam.
In principle we can get laser photons and initial electron beams with
proper helicity.
As a result, circular polarization of the back-Compton scattered
photon could be controlled.

Some SUSY particle production processes at the
$e\gamma$ and $\gamma\gamma$ colliders have already been discussed.
Particularly, analysis for the prosess
$\gamma\gamma\rightarrow\tilde{W}_{1}\tilde{W}_{1}$ have been given in
\cite{goto}, where only unpolarized initial beams are considered.
Here we focus our attention to the physical consequence of the initial
beam polarization.

Formulae for the polarized differential cross
sections of our process are obtained as follows ;
$$\frac{d\hat{\sigma}}{d\cos\theta}[\gamma_\pm\gamma_\pm\rightarrow\tilde{W}_i^+
\tilde{W}_i^-]=\frac{4\pi\alpha^2}{\hat{s}
(1-\hat{\beta}^2\cos^2\theta)^2}\hat
{\beta}(1-\hat{\beta}^4),$$
\begin{equation}\frac{d\hat{\sigma}}{d\cos\theta}
[\gamma_\pm\gamma_\mp\rightarrow
\tilde{W}_i^+\tilde{W}_i^-]=\frac{4\pi\alpha^2}{\hat{s}
(1-\hat{\beta}^2\cos^2\theta)^2}
\hat{\beta}^3\sin^2\theta(2-\hat{\beta}^2\sin^2\theta),
\end{equation}
where $\hat{s}\equiv s_{\gamma\gamma}$ and
$\hat{\beta}\equiv\sqrt{1-4m_{\tilde{W}_i}^2/\hat{s}}$.
Total cross sections are given by
$$\hat{\sigma}[\gamma_{\pm}\gamma_\pm\rightarrow
\tilde{W}_i^+\tilde{W}_i^-]=\frac
{2\pi\alpha^2}{\hat{s}}\left[2\hat{\beta}(1+\hat{\beta}^2)
+(1-\hat{\beta}^4)\ln
\frac{1+\hat{\beta}}{1-\hat{\beta}}\right],$$
\begin{equation}\hat{\sigma}[\gamma_\pm\gamma_\mp\rightarrow
\tilde{W}_i^+\tilde{W}_i^-]
=\frac{2\pi\alpha^2}{\hat{s}}\left[-2\hat{\beta}
(5-\hat{\beta}^2)+(5-\hat{\beta}^4)\ln
\frac{1+\hat{\beta}}{1-\hat{\beta}}\right],\end{equation}
$$\hat{\sigma}[{\rm unpol}]=\frac{2\pi\alpha^2}{\hat{s}}
\left[-2\hat{\beta}(2-\hat{\beta}^2)+
(3-\hat{\beta}^4)\ln\frac{1+\hat{\beta}}{1-\hat{\beta}}\right]. $$
Total enarge dependece of
each sub-process cross section $\hat{\sigma}$ is shown in Fig.1,where we take
 $m_{\tilde{W}}=100$ GeV.
When $\sqrt{\hat{s}}\displaystyle{\mathop{>}_{\sim}}{400}$ GeV,
$\hat{\sigma}(\pm,\pm)$ is  larger than any other
sub-process cross section. Since ${\hat{\sigma}}(\pm,\pm)$ has the $s$ wave
contribution. On the other hand, ${\hat{\sigma}}(\pm,\mp)$
becomes a most predominant for
$\sqrt{\hat{s}}\displaystyle{\mathop{<}_{\sim}}{400}$ GeV.

 $\hat{\sigma}(unpol)$ takes an average of
$\hat{\sigma}(\pm,\pm)$ and $\hat{\sigma}(\pm,\mp)$,where
$\hat{\sigma}(\pm,\pm)
\equiv\hat{\sigma}[\gamma_\pm\gamma_\pm\rightarrow\tilde{W}_i^+\tilde{W}_i^-]$,
etc.

Only arbitrary SUSY parameter which appeares in Eqs.(1) and (2)
is the chargino mass. Since our process is a pure SUSY QED process,
we can get results for both charginos
$\tilde{W}_1$ and $\tilde{W}_2$ with the formulae Eqs.(1) and (2).
For completeness, we also give the formulae for
$\gamma\gamma\rightarrow W^+W^-$,
which will be
needed in the discussion of background suppression;
$$\frac{d\hat{\sigma}}{d\cos\theta}[\gamma_\pm\gamma_\pm\rightarrow
W^+W^-]=\frac{2\pi\alpha^2}{\hat{s}(1-\hat{\beta'}^2
\cos^2\theta)^2}\hat{\beta}'
(3+10\hat{\beta'}^2+3\hat{\beta'}^4), $$
\begin{eqnarray}
&&\frac{d\hat{\sigma}}{d\cos\theta}[\gamma_\pm\gamma_\mp\rightarrow
W^+W^-]  \\
&& \quad =\frac{2\pi\alpha^2}{\hat{s}(1-\hat{\beta'}^2\cos^2\theta)^2}
\hat{\beta}'
[16-16\hat{\beta'}^2+3\hat{\beta'}^4+2\hat{\beta'}^2(8-3\hat{\beta'}^2)
\cos^2\theta+3\hat{\beta'}^4\cos^4\theta], \nonumber
\end{eqnarray}
where $\hat{\beta}'\equiv\sqrt{1-4m_W^2/\hat{s}}$.

The polarized cross section in Fig.1 for the sub-process
$\gamma\gamma\rightarrow$X is expressed as
\begin{eqnarray}
&&\hat{\sigma}(\xi_2(z_1),\xi_2(z_2))\nonumber \\
&& =\frac{1}{4}[(1+\xi_2(z_1))(1+\xi_2(z_2))\hat
{\sigma}[\gamma_+\gamma_+]+(1+\xi_2(z_1))(1-\xi_2(z_2))\hat{\sigma}
[\gamma_+\gamma_-] \nonumber \\
&&\quad +(1-\xi_2(z_1))(1+\xi_2(z_2))\hat{\sigma}[\gamma_-\gamma_+]
+(1-\xi_2(z_1))(1-\xi_2(z_2))\hat{\sigma}[\gamma_-\gamma_-]].
\end{eqnarray}
If the incident gamma beams were monochromatic, the formulae (2), (3)
and (4) would give just the cross section. However, since each gamma
is the secondary beam, the experimental cross section are obtained by
folding the sub-process cross section $\hat{\sigma}$ with the photon energy
spectra $ D_{\gamma/e}(z_i)$.
Experimental cross section $\sigma$ is given by
\begin{equation}
\sigma=\int_{0}^{z^{max}_{1}}\int_{0}^{z^{max}_2}dz_{1}dz_{2}D_{\gamma/e}
(z_1)D_{\gamma/e}(z_2)\hat{\sigma}(z_1,z_2).
\end{equation}
Here $z_i$ (i=1, 2) denotes the energy fraction of each
high energy gamma beam ;
$$z_i=\frac{E_{\gamma_i}}{E_e}. $$
In the following we take the upper limit on $z_i$ as
$z_i < 0.83$, which guarantees a good conversion efficiency
in the backward Compton scattering \cite{gg}.

Next we show the numerical results for the experimental cross section
Eq.(5).
The chargino mass dependence of the polarized and the unpolarized cross
section is shown in Fig.2, where we set $\sqrt{s}_{ee}$=1TeV.
We also plotted the total cross section for
$e^{+}e^{-}\rightarrow\tilde{W}_{1}^{+}\tilde{W}_{1}^{-}$ in Fig.2,
where we take $m_{\tilde{\nu}}$=500 GeV and consider two extreme cases ;
a) the lighter chargino is almost Wino ($\tilde{W}$) and
b) almost Higgsino ($\tilde{h}$).
First, we find that the polarized initial beams could enhance
the cross section for $m_{{\tilde{W}}_1}$ ${\displaystyle \mathop{>}_{\sim}}$
250 GeV.
Second, polarized $\gamma\gamma$ cross section dominates over the $e^+e^-$
one for $m_{{\tilde{W}}_1}$
${\displaystyle{\mathop{<}_{\sim}}}$
350 GeV. It should be noted that ${\tilde{W}_1}$ $=$ ${\tilde{h}}$ case
gives the maximum value of
$\sigma (e^{+}e^{-}\rightarrow\tilde{W}_{1}^{+}\tilde{W}_{1}^{-})$.
Third, we could know the mass of chargino in terms of
a measurement of the total cross section
$\sigma (\gamma\gamma\rightarrow\tilde{W}_{1}^{+}\tilde{W}_{1}^{-})$
only.
This is because the cross section depends on $m_{{\tilde{W}}_1}$
but not on any other arbitrary SUSY parameters as mentioned above.
On the other hand,
$\sigma (e^{+}e^{-}\rightarrow\tilde{W}_{1}^{+}\tilde{W}_{1}^{-})$
depends not only $m_{{\tilde{W}}_1}$ but also on
the chargino mixing angles and the mass of sneutrino.

Now we should discuss the experimental signature and the background.
For simplicity, we consider the case,
$m_{W}$ $<$ $m_{\tilde{W}_{1}}$ $<$ $m_{\tilde{f}}$.
In this case the chargino will dominantly decay into $W\tilde{Z}_{1}$,
where $\tilde{Z}_{1}$ denotes the lightest neutralino
and experimental signatures of our process would be
$W$-boson pair plus large missing energies.
Therefore, the most serious background will be
$\gamma\gamma\rightarrow{W}^{+}{W}^-$.
We have already given the formula Eq.(3) for the
$\gamma\gamma\rightarrow{W}^{+}{W}^{-}$ cross section.
Since the $W$-bosons come from the background process are emitted to the
back-to-back in the initial $\gamma\gamma$ CMS,
a cut on the acoplanarity $\phi_{acop}$ of the
$W$-boson pair will be effective to suppress the background processes.
In Fig.3 we show the transvers momentum $P^{q,q}_{T}$
distribution of expected number of events, where we impose a cut
on acoplanarity, $\phi^{qq}_{acop} > 90^{\circ}$,where $q^{,}$s denote the
quarks originated from the hadronic decays of the $W$-bosons.
Here we take $\sqrt{s}=\sqrt{s}_{ee}$=1TeV,
$m_{\tilde{W}_{1}}$=300 GeV, $m_{\tilde{z}_{1}}$=150 GeV
and the luminosity $L=1\mbox{fb}^{-1}$.
We see that if we take the polarization
$(\lambda_{1,2}.P^{c}_{1,2})=(+\frac{1}{2},-1)$,
the signal events could be distinguished from the background.

We have investigated the chargino production and focused our
attention to the physical consequence of the initial beam polarization.
It has been shown that appropriate beam polarization could be useful to
enhance the cross section for the chargino with the mass
$m_{\tilde{W_{i}}}\displaystyle{\mathop{<}_{\sim}}{0.4}\sqrt{s}$.
We have explicitly shown that the most serious $WW$ background can be
suppressed by the cuts on the acoplanarity and the choice of polarization.
Another good property of the process is the simple
dependence of the cross section
on the arbitrary SUSY parameters.
This could be enable us to measure the mass of chargino and
in turn to check the GUT relations among the gaugino masses.

\vskip 20pt

\begin{flushleft}
{\Large{\bf Acknowledgements}}
\end{flushleft}
Part of numerical calculations have been performed on FACOM M780
at INS.
  \newpage

\newpage
{\bf\huge Figure Captions}
\vskip 20pt

Figure 1: $\sqrt{s}$ dependence of total sub-process cross section for
each photon polarization $(\xi_{2}(z_1),\xi_{2}(z_2))$.
We take $\sqrt{s_{\gamma\gamma}}=$1TeV.

Figure 2: Chargino mass dependece of total cross sections for each
initial beam polarizatin $(\lambda_1,P_1^c)$ and $(\lambda_2,P_2^c)$ .
We take $\sqrt{s_{ee}}=$1TeV.

Figure 3: Monte-Carlo event generation for
$\gamma\gamma\rightarrow\tilde{W_1^+}\tilde{W_1^-}$
and $\gamma\gamma\rightarrow{W^+W^-}$
with cut $\phi_{acop}>90^\circ$ .
We take $\sqrt{s_{ee}}=$1TeV, $m_{\tilde{W_1}}=$300GeV,
$m_{\tilde{Z_1}}=$150GeV, $(\lambda_{1,2},P_{1,2}^c)
=(\frac{1}{2},-1)$ and the luminosity
$L=\mbox{1fb}^{-1}$.

\end{document}